# A complex network theory approach for the spatial distribution of fire breaks in heterogeneous forest landscapes for the control of wildland fires

Lucia Russo[a], Paola Russo[b], Constantinos I. Siettos[c]

[a]Combustion Research Institute, Consiglio Nazionale delle Ricerche, Naples, Italy

[b] Department of Chemical Engineering, Materials and Environment, Sapienza University of Rome, Rome, Italy

[c]School of Applied Mathematics and Physical Sciences, National Technical University of Athens, Greece

**Abstract.** Based on complex network theory, we propose a computational methodology that addresses the spatial distribution of fuel breaks for the inhibition of the spread and size of wildland fires on heterogeneous landscapes. This is a two-tire approach where the dynamics of fire spread are modeled as a random Markov field process on a directed network whose edge weights, are provided by a state-of-the-art cellular automata model that integrates detailed GIS, landscape and meteorological data. Within this framework, the spatial distribution of fuel breaks is reduced to the problem of finding the network nodes among which the fire spreads faster, thus their removal favours the inhibition of the fire propagation. Here this is accomplished exploiting the information centrality statistics. We illustrate the proposed approach through (a) an artificial forest of randomly distributed density of vegetation, and (b) a real-world case concerning the island of Rhodes in Greece whose a major part of its forest burned in 2008. Simulation results show that the methodology outperforms significantly the benchmark tactic of random distribution of fuel breaks.

**Keywords:** Wildland Fires; Hazard Management; Fuel Breaks; Cellular Automata; Markov Field; Complex Networks

# INTRODUCTION

Efficient wildland fire-prevention is one of the most challenging and important problems in ecology [1] to reduce irreversible environmental and socio-economic damages. Fuel management treatments have been extensively applied at local scale, but they have a limited influence on the evolution of wildfires at the landscape scale [2, 3]. At this scale level, experimental work is prohibitory, and the majority of previous studies on the spatial distribution of fuel management activities have been mostly theoretical. Observations on real wildland fire cases have evidenced that fire size and severity can be mitigated by proper design of treatments such as fuel segmentation and prescribed burning [3-6]. Hence, organizations like the USDA Forest Service, in their fuel managements plans, consider spatial fragmentation of the fuel crucial [7]. The problem is how to spatially distribute the fuel management activities across the landscape. Towards this aim, there is no doubt that the systematic -in terms of mathematical modeling and analysis – quantification of the fire spread dynamics is of outmost importance. Factors such as weather/climate conditions (wind field, air humidity and temperature), characteristics of the distributed local fuel (type and structure of the vegetation, moisture and density), landscape/earth characteristics (slope, fragmentation and natural barriers) as well as fire-suppression tactics are considered to be key elements [3, 8-14]. Within this context, methods coming from mathematical modeling and analysis of complex systems can thus help enhancing our efficiency in designing better control policies. Clearly, the core of the approach is the adoption of accurate and efficient simulation models. At small scale, previous theoretical studies on the spatial distribution of fuel managements activities have shown that a random or compartmented distribution of fuel management activities reduce the spread rate of the fire when dealing with large portions of the landscape [15, 16]. On the other hand, regular patterns like parallel stripes work effectively just if the fire moves perpendicular to the stripes [17]. Although these approaches are very promising, relatively fewer studies have faced the problem at the real-world landscape level using mathematical simulation techniques [2, 18-21]. In [2], both operational and fuel reduction management tactics are optimized on the landscape level, where spatial patterns of fuel management activities were scheduled using a heuristic optimization algorithm that was implemented around FARSITE [22]. In [18], several management activities, namely prescribed burning, thinning and timber harvesting are evaluated and compared across four landscapes with increasing fuel content. The distribution of activities is achieved using optimization which was wrapped around SIMMPPLE [23]. In [19], fire mitigation through the implementation of fuel breaks was simulated on in the Mara͂o mountain range, NW Portugal. The authors used FARSITE [22] to simulate the propagation of the fire spread. In [20], several intensities of fuel treating stands considering also spatial density and residential priorities on a 16,000 ha study area in Oregon, US were evaluated with respect to forest health and ecological restoration. Wildland fire spread was simulated using the minimum travel time (MTT) fire spread algorithm [24]. In [21], the authors compared the relative effectiveness of thinning, thinning and burning, and burning on simulated stand-scale fire behavior as well as the effectiveness of three different arrangement of treatments, namely a random, defensible fuel profile zones (DFPZ) [25], and strategically placed area treatments (SPLATs) [17] on a simulated area of Southern Cascade range, California.

Here, we focus on the problem of the systematic spatial distribution of fire breaks as realized by vegetation cutting for the hazard management of wildland fires. Except from the inhibition of fire spread, vegetation cutting is commonly used to restore longleaf pine forests, enhance the growth rate of young trees reducing midstorey and shrub layers and at the same time can be also used for commercial thinning of woodlands. Towards this aim, motivated by the arsenal that complex network theory offers, we propose a computational two-tier framework, where the fire spread is being modeled to evolve through a (lattice) weighted directed network whose edge weights are the state transition probabilities of a state-of-the-art spatio-temporal Markov Cellular Automata (CA) process [26, 27]. The particular CA model incorporates detailed GIS, landscape and meteorological data and has been proved to be robust and efficient in predicting the fire spreading behaviour in several real-world cases [26-29]. Then, the problem of the spatial distribution of fire breaks is reduced to the problem of finding the group of nodes through which the fire spreads faster. Actually this problem is closely and straightforwardly related to the analysis of information flow on networks. Here, for the detection of information hubs (fire spread hubs) we make use of the Bonacich information centrality criterion [30]. Our approach is illustrated through two examples: (a) an artificial forest of randomly distributed density of vegetation, and (b) a real-world case concerning the island of Rhodes in Greece whose a major part of its forest burned in 2008. Simulation results over an ensemble of lattice realizations and runs show that the proposed approach appears very promising as it produces statistically significant better outcomes when compared to the benchmark policy of the random distribution approach.

## MATERIALS AND METHODS

The proposed methodology aims at the systematic design of the spatial distribution of fire breaks for the inhibition of the propagation of wildland fires in heterogeneous forest landscapes. Fire propagation is being modeled as a random walk on a network (lattice) with directed weighted links; a Cellular Automata model defines the transition matrix $P$ of the random walk processes [31]. Then, the spatial distribution of fire breaks is reduced to a flow information problem: that of finding a partition of network nodes that if removed from the network the flow information (here the propagation of fire through the network) slows down. These nodes are the nodes through which the information flows faster within the network. The terrain is tessellated into a number of small patches whose shape and size depends on the level of precision with respect to the spatial directionality (squares, hexagons etc.). Hence, the terrain is transformed into a lattice network, say, $G(V, E)$, where $V = \{v_k\}, k = 1, 2, ..., N$ is the set of nodes (the terrain patches), and $E$ is the set of edges (links) between neighbor nodes. Each edge $e_{v_k \to v_l}$ is directed and its weight is determined by the states of the nodes $v_k, v_l \in V$ associated with it as it will be explained below. Here, the system state over the set of the nodes (edges) is denoted by $S(V)$ ($S(E)$), where $S(v_k) \equiv s_i(v_k) = \{s_{ik}\}, i = 1, 2, ..., M$ is the set of $M$ states of node $v_k$. The states can be continuous ($s_{ik}(\cdot) \in R$) or discrete ($s_{ik}(\cdot) \in Z$) and represent both terrain and forestry characteristics such as elevation, type and density of vegetation, moisture content, vegetation height, etc.; in addition to the landscape characteristics they also include a component associated to the susceptible-burning-burned states, say, $s_{1k} = 0$ if $v_k$ is still susceptible to burning, $s_{1k} = 1$, if $v_k$ is burning and $s_{1k} = -1$, if $v_k$ has been already burned. Let us also denote the neighbourhood of $v_k$ as $\Re_{v_k}$. For a von Neumann lattice, the neighbourhood of each node $v_k$ is defined by a set of the surrounding 8 nodes (Moore-neighbourhood). Thus, system dynamics can be represented as a spatio-temporal Markov process of order two in space and order one in time (*STMC(1,2)*) with transition probabilities:

$$p(s_{1k}(t+1)=-1 \mid s_{1k}(t)=1)=1 \qquad (1)$$

(the above relation implies that a node $v_k$ that is burning at time $t$ will be burned down at the next time step),

and

$$p(s_{1k}(t+1)=1 \mid s_{1l}(t)=1)=p_{v_l \to v_k} \qquad (2)$$

where, $v_l \in \Re_{v_k}$ and $p_{v_l \to v_k} = f(s_{ik}, s_{il})$ is the probability that the fire spreads from $v_l \in \Re_{v_k}$ to $v_k$ and depends only upon all the other states of the $v_k$, $s_{ik, i=2,3,...,M}$, and of all the other states of its neighbors, $s_{i\Re_{v_k}, i=2,3,...,M}$. In the general case, $p_{v_l \to v_k} \neq p_{v_k \to v_l}$, so the network is directed. The transition probability $p_{v_l \to v_k}$ defines the directed weight of the spatial connection: $s(e_{v_l \to v_k}) \equiv s_{e_{v_l \to v_k}} = p_{v_l \to v_k}$. Self-contacts are not allowed, i.e. $s_{e_{v_k \to v_k}} = 0$. Hence, the corresponding adjacency matrix, $A$, is a non-symmetric matrix of size $N \times N$ whose elements $a_{lk}$ are the states of the links $s_{e_{v_l \to v_k}}$. Under the above formalism the system dynamics can be presented as a second order directed Markov field.

Having computed the adjacency matrix, centrality statistics can be then used for the identification of nodes that contribute more to the information (fire spread) flow through the network. Thus nodes with a high value of centrality are connected with other nodes by relatively short paths and so they are central to the information flow (and therefore the fire spread) through the network. Under this view, we rank cells according to their centrality values from higher to lower values and we remove (i.e. cut the vegetation in the corresponding land patches) the first, say, $N_e \subset N$, of them.

Several measures have been proposed for the assessment of the information centrality in networks such the Betweenness centrality (BC), the closeness centrality (CC), the degree centrality (DC), the Bonacich centrality (BC) and the Eigen-centrality (EC) [30, 32].

The BC of node $v_k$ is defined as:

$$BC_{v_k} = \sum_{l \neq k \neq m} \frac{n_{v_l v_m}^{v_k}}{n_{v_l v_m}}. \qquad (3)$$

$n_{v_l v_m}^{v_k}$ is the number of shortest paths between nodes $v_l$ and $v_m$ passing from $v_k$, and $n_{v_l v_m}$ is the number of the shortest paths between $v_l$ and $v_m$.

The CC of node $v_k$ is defined as the inverse of the sum of geodesic distances (i.e. the shortest paths) from node $v_k$ to all other nodes in the network:

$$CC_{v_k} = \left( \sum_{m=1, k \neq m}^{N} d_{v_k v_m} \right)^{-1}. \tag{4}$$

$d_{v_k v_m}$ is the geodesic distance from $v_l$ to $v_m$.

The Eigencentrality of node $v_k$ corresponds to the *k-th* component of the eigenvector related to the largest eigenvalue of the adjacency matrix $A$. For very large scale directed weighted networks where $A$ is asymmetric one can employ Arnoldi's iterative method for extracting a low dimensional upper Hessenberg matrix whose eigenvalues of provide approximations of the outermost spectrum of the full matrix [33]. However for directed graphs Eigencentrality may not produce meaningful results [30]. An extension of Eigencentrality to directed weighted graphs comes from the Bonacich measure defined as the *k-th* component of [30]:

$$x = \left( I - \frac{\beta}{\lambda_{max}} A \right)^{-1} e, \tag{5}$$

where $e$ is a vector of ones and $\lambda_{max}$ is the largest eigenvalue of $A$ (that can be computed through the Arnoldi eigensolver). Note that for $\beta = 0$, the above expression reduces to the degree centrality, while for $\beta = 1$ it reduces to the standard Eigencentrality.

To illustrate the efficiency for each one of the above measures we first considered a simplistic case of a square von-Neumann Lattice with Moore Neighborhood and periodic boundary conditions, where the landscape is flat, the type of vegetation is homogenous, i.e. there is just one type of vegetation (e.g. pine trees) and the density of vegetation varies in continuously from

1 (corresponding to empty/burned cells) to 0 (corresponding to very dense vegetation). The state of a node $v_k$ at time $t$ is represented $s_{2k}(t) \in [0\ 1]$. Then, dynamics advance from time $t$ to time $t+1$ for all nodes simultaneously according to the following rules: $v_l \in \Re_{v_k}$

*Rule 1:* IF $s_{1k}(t) = 1$ THEN $s_{1k}(t+1) = -1$.

This rule implies that a burning node at the current time step will be burned down at the next time step.

*Rule 2:* IF $s_{2k}(t) \in [0\ 1)$ THEN IF $s_{v_l}(t) = 1$ THEN $s_{1k}(t+1) = 1$ with probability $p_b = 1 - s_{2k}(t)$, $v_l \in \Re_{v_k}$.

This rule implies that if a node contains fuel then if there is a burning node at its neighborhood, then it will catch fire at the next time step with probability that is proportional to the density of vegetation.

In the random field Markov process framework the above CA rules can be written compactly as:

$$p(s_{1k}(t+1) = -1 \mid s_{2k}(t) = 1) = 1, \tag{6}$$

and

$$p(s_{1k}(t+1) = 1 \mid s_{1\Re_{v_k}(t)} = 1) = 1 - s_{2k}(t). \tag{7}$$

The approach is tested against the random distribution of fire breaks benchmark with respect to the hazard intensity as a function of the density of fire breaks defined as:

$$R(d_f) = \frac{1}{N_r} \sum_{i=1}^{N_r} \frac{N_b(i)}{N_v}. \tag{8}$$

$d_f = \frac{N_e}{N_v}$ is the number of fire breaks nodes divided by the total number of nodes that contain flammable vegetation. $N_r$ denotes the number of simulations for a given initial condition, $N_b$ denotes the total number of burned cells and $N_v$ denotes the total number of nodes (cells) that contain vegetation within the area of interest. The above index can be interpreted as the maximum likelihood ratio of burned area.

Significant differences for a given $d_f$ was computed by implementing the non-parametric Wilcoxon test on the $N_r$ outcomes of the simulations with a threshold set at $a = 0.01$.

The key and crucial element for the extension of the above approach to real-world cases is the computation of reliable transition probabilities. When dealing with real-world problems there is a basic question to be answered: how much could we trust the outcomes for a real situation? State-of-the-art models are-as all models- just approximations of the real system they represent. Due to the inherent extraordinarily complexity of the problem, they are built with incomplete knowledge and for that reason they are flashing a "note of caution" on parameter and rule inaccuracies. Toward this aim we employed a state-of-the-art CA model that has been developed over the past few years and has been shown to be robust and efficient in predicting the fire spreading behaviour in several real-world cases [26-29]. In particular, keeping the same values of the model parameters as the ones found by optimization for the case of Spetses island in Greece in 1990 [26] the model predicted quite well the dynamics of the large scale wildfire that occurred in the mountain of Parnitha, Greece in 2007 and was one of the most catastrophic fire incidents in

Greece over the last 50 years [27] and that of of wildland fire which devastated Rhodes in 2008 [28].

The proposed CA model takes into account major macrosopic factors that determine the course of the fire like the vegetation density, type, height moisture air content, the wind field and detailed GIS data. For each node $v_k$ we consider the following states:

$$\{s_{2k}, s_{3k}, s_{4k}\} = \{\text{impact of the type of vegetation, impact of the density of vegetation, node's elevation}\}$$

In a nutshell, fire dynamics (considering no wind conditions) are propagated from a node $v_l$ to its neighbors according to (2), where:

$$p_{v_l \to v_k} = p_0 (1 + s_{2k})(1 + s_{3k}) f(s_{4k}, s_{4l}). \qquad (9)$$

$p_0$ is a nominal probability of fire spread under no wind, flat terrain and certain density and type of vegetation and it is calculated from experimental data. The type and the density of vegetation in the area are split into a number of discrete categories. In particular, the type and the density of vegetation in the area are split into a number of discrete categories. For the case under study (Rhodes island), the type of vegetation has been clustered into three categories (agricultural areas, shrubs and pine trees), while the density of vegetation has been scaled into three categories (sparse, normal, dense). $f(s_{4k}, s_{4l})$ denotes the effect of the slope between nodes $v_l$ and $v_k$ and is calculated via:

$$f(s_{4k}, s_{4l}) = exp(a\theta_s). \qquad (10a)$$

$\theta_s$ is the slope angle between $v_l$ and $v_k$ and $a$ is a constant that can be adjusted from experimental data. For a square grid, the slope angle is calculated in a different way depending on whether the two neighboring nodes are adjacent or diagonal to the burning node. More specifically for adjacent node the slope angle reads:

$$\theta_s = \tan^{-1}\left(\frac{s_{4l} - s_{4k}}{l}\right), \tag{10b}$$

where $l$ is the length of the square side. For diagonal nodes the formula becomes:

$$\theta_s = \tan^{-1}\left(\frac{s_{4l} - s_{4k}}{l\sqrt{2}}\right) \tag{10c}$$

The parameter values for the above parameters for the case of the Greek cases are given in table 1.

Finally, the weights of the links are inversed so that long walks count to slow spread directions while short walks count for fast spread directions.

# SIMULATION RESULTS

Simulation results were performed in Matlab programming environment.

*The toy problem*

For illustrating our approach we first considered the simple case of a random distribution of a single vegetation density on a flat terrain of a lattice of 50x50 cells with periodic boundary conditions. Random numbers were created by a uniform distribution in (0 1) using Matlab's function *rand*. We used $N_r = 100$ realizations (ensembles) of randomly generated forest and for each one of the realizations we created a corresponding distribution of fire breaks (randomly or network-based) and we run the simplistic CA model, until there were no burning cells. All simulations started by setting a fire at the center of the lattice. Fig. 1 depicts the diagram of $R(d_f)$ with respect to the percentage of fire breaks (empty cells), $d_f$, for the centrality measures described at the Methods and Material section. The results obtained with the random distribution tactic are also shown for comparison purposes. Eigencentrality fails in identifying an adequate partition of nodes, while the Bonacich measure for a wide range of β values gives equivalent results with the BC measure. Thus the shaded area shows the region where differences between the Bonacich and the BC-based are statistically significant from the random-based distribution. The comparison was made using the non-parametric Wilcoxon statistical test with a threshold set at $a = 0.01$. As it is clearly shown, the proposed approach based on the Bonacich and the BC criterion outperforms the random-based distribution of fire breaks. For the Bonacich and the BC-based approach there is a relatively sharp phase transition from high to low $R$ s around $d_f = 0.22$ while for the random-based distribution the phase transition occurs around $d_f = 0.32$.

*The real case of Rhodes Island, Greece*

We applied the proposed methodology to simulate the effect of the distribution of fire breaks in the south part of Rhodes Island Greece. Fig. 2 shows a stereoscopic image of Rhodes; the resulted burned area is also overlaid. We have chosen this case as in July 2008 Rhodes swept through a wildland fire that resulted to huge damage. The forest fire occurred on 22.07.2008 and broke out at 11.40 am at the Ag. Isidoros point, Municipality of Attaviros. The disastrous fire caused huge ecological disturbance like the incineration of thousands of Pinus brutia that was among the dominant species of threes in the burned area. Fig. 3 depicts a map of the vegetation density of the area under study. The cause of the fire was identified as negligence and the wind was NW-4-5 bf at the time of the eruption. The consequences of the fire were 13,240 ha of burnt area and an inestimable environmental disaster. Damage included also many infrastructures like one destroyed house and many store houses, cultivated areas, machinery, equipment and domestic animals. The forest fire was fought for five days with 1.230 fire fighters from the Fire Brigade Service, 1.000 people from the Greek armed forces, 200 volunteers, 75 fire fighting vehicles, 46 other vehicles, 10 aircrafts and 9 helicopters. The total suppression cost was estimated at about 16 mill. €. The fire evolved rapidly under conditions of heterogeneous mountainous landscapes, density and characteristics of the vegetation as well as significant meteorological changes. In order to demonstrate the appropriateness of the CA model we first simulated the event taking into account also the weather conditions in the area during the period of the incident. It should be noted that as the model is stochastic, simulations will generally result to different $N_b$s (number of burned cells). For this reason, we have chosen to run the model $N_r = 100$ times and depict all simulations by mapping the burning frequency of each cell in a gray-scale mode. Hence, for each cell we calculated the average (expected) number of burnings out of the 100 simulation runs. Then, we created the corresponding histogram of average number of burnings for the area under study using $2^4$ bins. Finally, we assigned - in descending order - $2^4$ levels of gray (from $2^8$ to $2^4$ with a discretization step of $2^4$) to the bins. The resulting "expected" map of relative burning frequencies is shown in Fig. 4a. For comparison purposes Fig. 4b illustrates the actual burned area overlaid on the vegetation map. As it is shown the "expected" burned area predicted by simulations is quite close to the real one. It is clear that the proposed model captures quite satisfactory the behavior of the fire, which is a rather challenging task taking into account the difficult heterogeneous terrain as well as the varying vegetation type and density. We then applied the proposed approach for the risk management of a particular area (defined by the rectangular shown in Fig. 4b) using the Bonacich measure with a β=0.5 (choosing different values of β resulted to equivalent results). Fig. 5 depicts the diagram of average $R(d_f)$ over $N_r = 100$ runs. The results obtained with the random distribution tactic are also shown for comparison purposes. The shaded area illustrates differences which are statistically significant computed by the non-parametric Wilcoxon criterion with a threshold set at $a = 0.01$. Maximum values of the burned area are also illustrated with bars.

As it is shown, the proposed approach results to significant hazard reduction when compared with the random distribution. In particular for $d_f = 0.14$ only the ~3.7% of the forest in the marked area is burned while the maximum value of $R(0.14)$ over the 100 runs is also very low ~5% (this value raised up to ~9% for $d_f = 0.16$). For the same level of vegetation cutting, the random distribution tactic resulted to ~23% burned forest with a maximum of ~65%. It is worth pointing out that when using the random distribution of fire breaks in order to achieve the same average value (over the 100 runs) of $R$ at $d_f = 0.14$ with the one obtained with the proposed approach, $d_f$ has to be increased to 19%. Furthermore, with the proposed approach for $d_f > 0.14$ the maximum value of burned forest over 100 runs is very low (below 10%), while with the random distribution tactic there is still a high probability that a large portion of the forest may be burned as the maximum value of $R$ in that case may exceed 40% (e.g. at $d_f = 0.17$). Fig. 6a (b) shows the location of fire breaks for the proposed (random) approach at the level of $d_f = 0.14$ (0.19). Interestingly, the results showed that for any practical means there are not significant differences in the selection of sparse, normal or dense patches between the proposed approach and the random distribution. With the proposed approach ~42% of the fire breaks are located in nodes with sparse density, ~31% in nodes with normal density and ~27% in nodes with dense density of vegetation. The values for the random distribution are ~42%, ~30% and ~28%, respectively. Hence with almost the same distribution of fire breaks as the level of density is concerned, the proposed approach succeeded to the total inhibition of the fire spread at an expense of <5% of the total forestry.

# DISCUSSION

Uncontrolled wildfires have been the cause of numerous irreversible environmental damages with serious negative ecological and socio-economic consequences including loss of human lives, flora and fauna bio-diversity and rare-species destruction, habitant fragmentation, floods, loss of timber harvest capability, economic loses in the tourism sector, air pollution and climate change. Hence, one of the most challenging problems in ecology revolves around the design and implementation of efficient wildland fire-prevention. The systematic -in terms of mathematical modelling and analysis – quantification of the fire spread dynamics is of outmost importance towards the risk assessment of a potential outbreak. However, due to the inherent complexity of such a phenomenon deploying at different time and space scales, risk assessment is far from simple. The computational approach proposed here is devoted to the design of the distribution of fire breaks with the aid of complex network theory and detailed Cellular Automata modeling. The proposed technique is based on the concept of centrality criterion, a key statistical measure for the evaluation of information flow through complex networks. The approach involves the construction of the adjacency matrix of the network, through which the fire propagates, whose elements are the strengths (weights) of the fire propagation. The CA-based model that was used to calculate the transition probabilities of fire spread among the land patches can deal with spatial heterogeneity in both the fuel and landscape characteristics can be coupled with Geographical Information Systems (GIS) and can take as input local meteorological data (even in real time). It has been proven to be robust and efficient in predicting the fire spreading dynamics both in space and time in several real-world large-scale wildland fires [26-29]. The proposed approach is here illustrated through (a) a simplistic lattice configuration which encompasses a single type of vegetation with randomly varying density and (b) a real world case, that is the wildfire occurred in the island of Rhodes Greece in 2008. Our approach is compared with the benchmark random based distribution of fuel breaks. It is shown that the proposed approach succeeds in inhibiting the spread of fires in heterogeneous landscapes at relatively low levels of vegetation cutting. The methodology can be combined with contemporary forestry management that makes use of either properly designed surface fires to restore longleaf pine ecosystems [34, 35] or low-intense tree cutting that can under certain restrictions enhance forest sustainability. It would also interesting as a future work to compare the efficiency of the approach with other proposed studies [2, 18-21] but this is beyond the scope of the current study.

To this end, we should point out that the implementation of such an approach (as other similar approaches) should not be considered as a black-box one. Additional important issues and restrictions related to the effects of tree cutting on the forestry ecosystem (e.g. such as its influence to biodiversity) have to be taken into account [20, 36-39].

# TABLES

**Table 1**. *Parameter values for the CA model*

| Parameter | Value | Effect of Density | $s_{3k}$ | Effect of Type | $s_{2k}$ |
|---|---|---|---|---|---|
| $p_o$ | 0.58 | Sparse | -0.4 | Agricultural | -0.3 |
| $a$ | 0.078 | Normal | 0 | Shrubs | 0 |
|  |  | Dense | 0.3 | Pines | 0.4 |

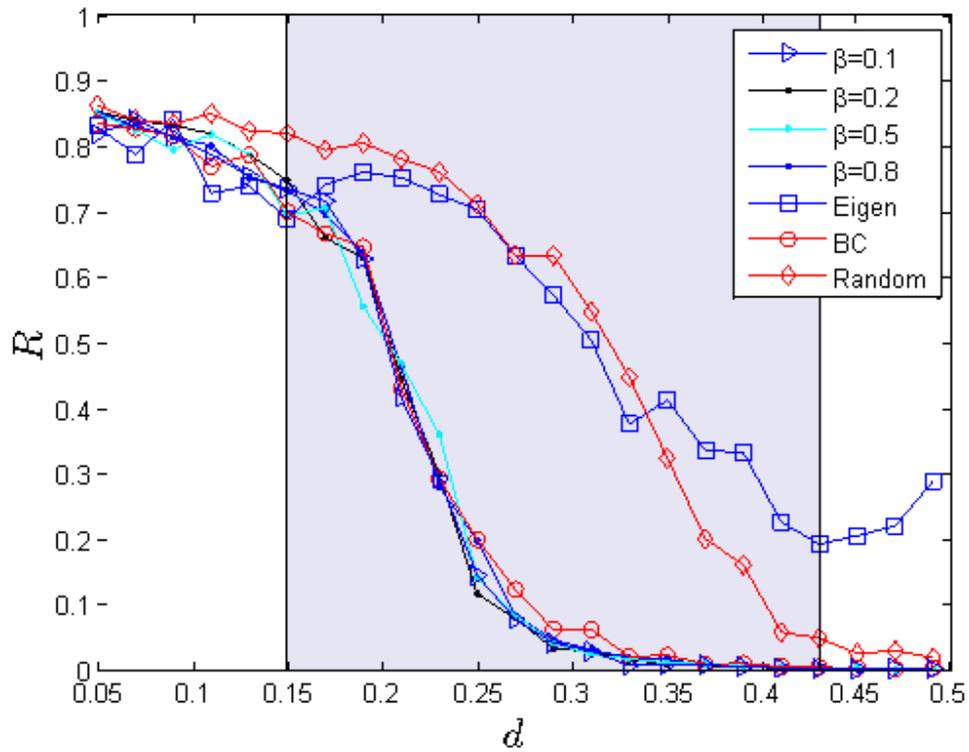

**Figure 1.** Maximum likelihood ratio of burned area $R(d_f)$ with respect to the density of fire breaks, $d_f$, as distributed with various centrality measures. The random approach is also depicted

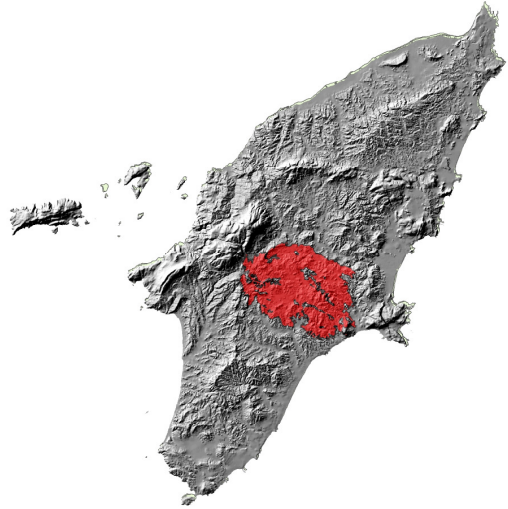

**Figure 2.** A stereoscopic image of Rhodes island Greece.

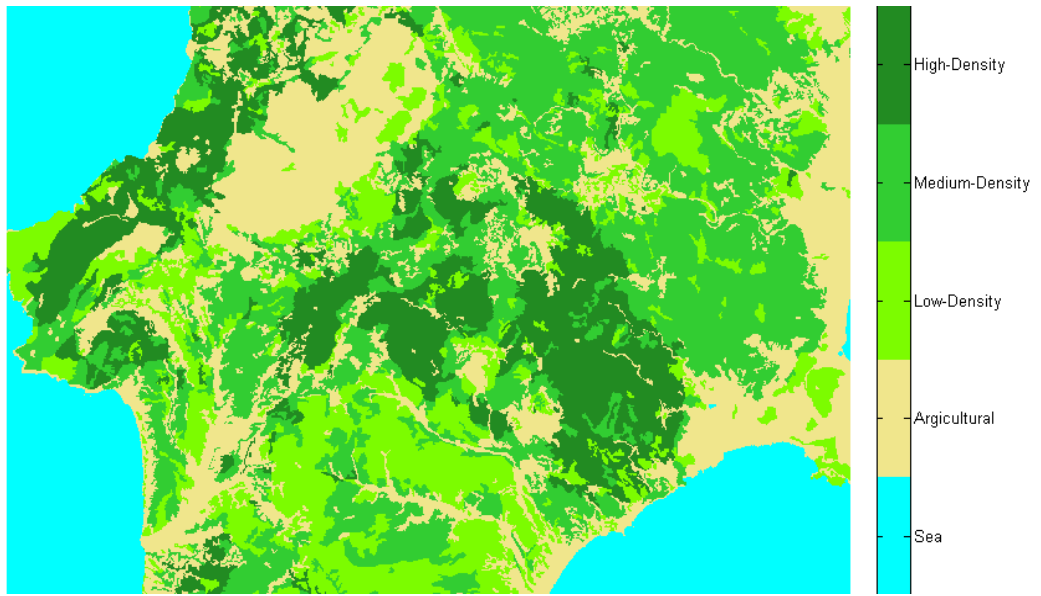

.**Figure 3**. Map of the vegetation density of the area under study in Rhodes island,Greece.

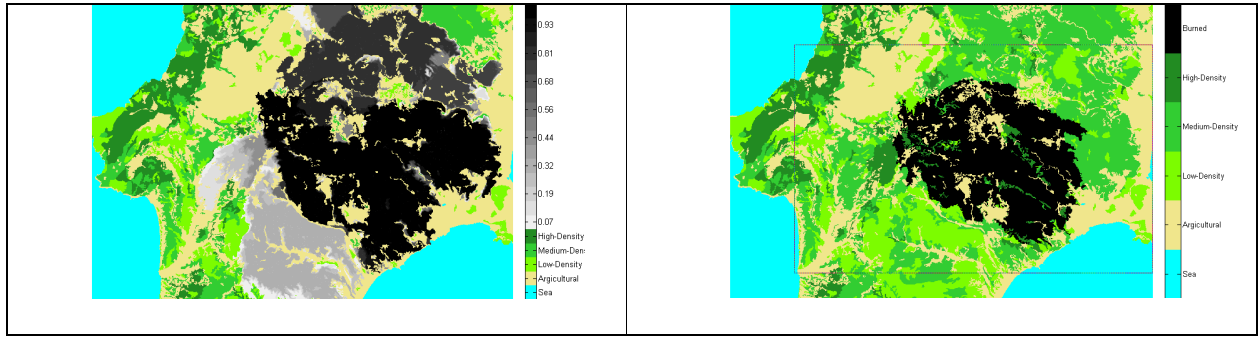

**Figure 4. (a)** The "expected" map of relative burning frequencies. This was constructed using the CA model. **(b)** Actual burned area overlaid on the vegetation map.

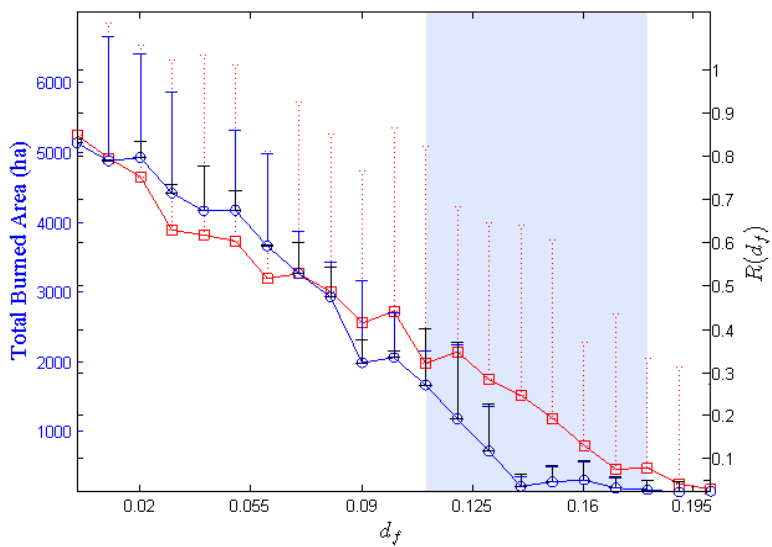

**Figure 5.** Maximum likelihood ratio of burned area, $R(d_f)$, over $N_r = 100$ runs (circles). The results obtained with the random distribution of fire breaks is also shown (squares). The shaded area statistically significant differences as computed by the non-parametric Wilcoxon criterion (at $a = 0.01$). Maximum values of the burned area are also illustrated with bars.

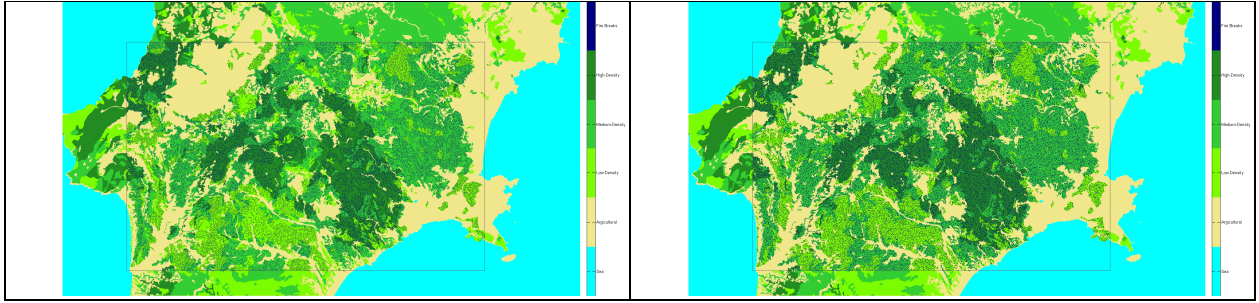

**Figure 6. (a)** Distribution of fire breaks using the proposed approach for $d_f = 0.14$. For this level of the density of fire breaks the expected maximum likelihood of burned area was ~3.7%, while the maximum value of $R(0.14)$ over 100 runs was ~5% (see Figure 5). **(b)** Random distribution of fire breaks for $d_f = 0.19$. For this level of the density of fire breaks the expected maximum likelihood of burned area was ~8%, while the maximum value of $R(0.19)$ over 100 runs was ~33% (see Figure 5).